\documentclass[%
 superscriptaddress,
 amsmath,amssymb,
]{revtex4-1}

\usepackage{amsmath}
\usepackage{amssymb}
\usepackage{url}
\usepackage{verbatim}
\usepackage[utf8]{inputenc}

\begin{document}

\title{The Nested\_fit data analysis program} 

\author{M. Trassinelli}
\email[]{martino.trassinelli@insp.jussieu.fr}
\affiliation {Institut des NanoSciences de Paris, INSP, CNRS, Sorbonne Université, F-75005 Paris, France}

\date{\today}

\abstract{
We present here \texttt{Nested\_fit}, a Bayesian data analysis code developed for investigations of atomic spectra and other physical data.
It is based on the nested sampling algorithm with the implementation of an upgraded lawn mower robot method for finding new live points. 
For a given data set and a chosen model, the program provides the Bayesian evidence,  for the comparison of different hypotheses/models, and the different parameter probability distributions. 
A large database of spectral profiles is already available (Gaussian, Lorentz, Voigt, Log-normal, etc.) and additional ones can easily added.
It is written in Fortran, for an optimized parallel computation, and it is accompanied by a Python library for the results visualization.
}

\maketitle

\section{Introduction}

\texttt{Nested\_fit} is a general purpose parallelized data analysis code for the evaluation of \textit{Bayesian evidence} and parameter probability distributions for given data sets and modeling function.
The computation of the Bayesian evidence is based on the nested sampling algorithm \cite{Skilling2004,Sivia,Skilling2006}, for the integration of the likelihood function over the parameter space.
This integration is obtained reducing the $J$-dimensional volume (where $J$ is the number of parameters) in a one-dimensional integral by a clever exploration of the parameter space.
In \texttt{Nested\_fit}, this exploration is obtained with a search algorithm for new parameter values called \textit{lawn mower robot}, which has been initially developed by L.~Simons \cite{Theisen2013} and modified here for a better exploration of multimodal problems.

\texttt{Nested\_fit} has been developed over the past years to analyze several sets of experimental data from, mainly, atomic physics experiments.
For this reason, it has some special feature well adapted to the analysis of atomic spectra as specific line profiles, possibility to study correlated spectra at the same time, eg. background and signal-plus-background spectra, and with a likelihood function built considering a Poisson statistics per each channel, well adapted to low-statistics data.


In the next section we will describe the general structure and feautres of \texttt{Nested\_fit}.
In Sec.~\ref{sec:nested} we shortly introduce the basic concepts of Bayesian model comparison and the nested sampling method.
The specific algorithm for the parameter space exploration for the nested sampling  is presented in details in Sec.~\ref{sec:lawn-mower}.
An example of application of  \texttt{Nested\_fit} is presented in Sec.~\ref{sec:go} for the analysis of single two-body electron capture ion decay.
A conclusive section will end the article, where recent application of \texttt{Nested\_fit} to different atomic physics analysis are mentioned.

\section{General structure of the program}

\begin{figure}
\includegraphics[width=\textwidth]{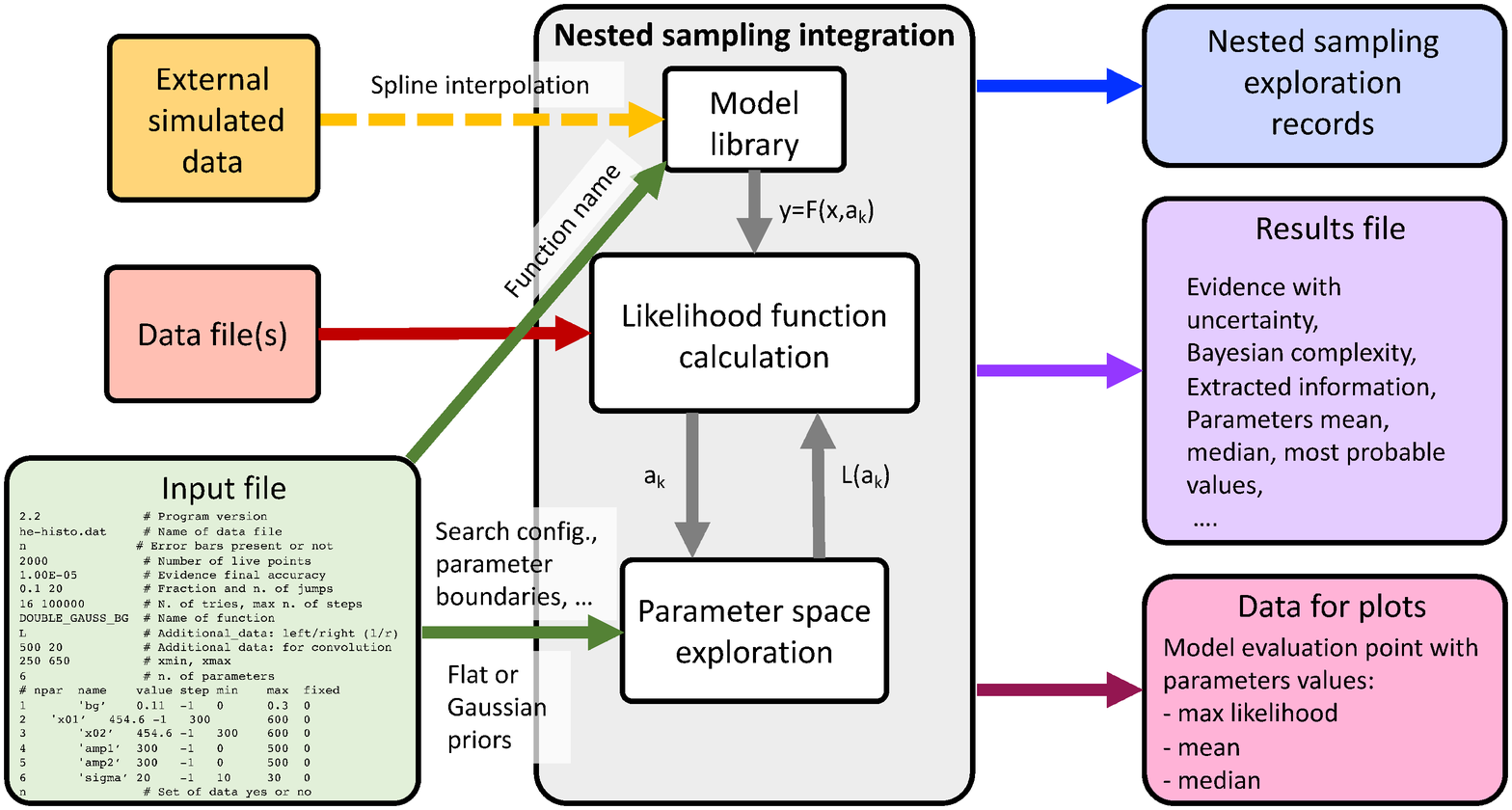}
\caption{Scheme of the \texttt{Nested\_fit} program.}
\label{fig:scheme}
\end{figure} 

The general structure of the program is represented in Fig.~\ref{fig:scheme}.
The main input files are two: the file  \texttt{nf\_input.dat}, where all computation input parameters are included, and the data file, which name is indicated in the parameter input file.
The function name in the input file indicates the model to be used for the calculation of the likelihood function. 
Several functions are already defined in the function library for modelling spectral lines: Gaussian, Log-normal, Lorentzian, Voigt (Gaussian and Lorentzian convolution), Gaussian convoluted with an exponential (for asymmetric peaks), etc.
Additional functions can be easily defined by the users in the dedicated routine (\texttt{USERFNC}).
Differently from the version presented in Ref.~\cite{Trassinelli2017b} (V.~0.7), in the new version discussed here (V.~2.2) non-analytical or simulated profile models can be implemented.
In this case, one or more additional files have to be provided by the users. 
These external data, which can have some noise like the case of simulated data, are interpolated by B-splines using FITPACK routines \cite{Dierckx}.
The B-spline parameters are stored and used as profile/model with the total amplitude and a possible offset as free parameters.
An additional feature of this new program version, is the possibility to analyze data with error bars. 
This option has to be indicated in the input file.


Several data sets can be analyzed at the same time by selecting the option ``set of data: YES''. 
This is particularly important for the correct study of physically correlated spectra at the same time, eg. background and signal-plus-background spectra.
This is done using a global user-defined function with common parameters of specific models for each spectrum.
In the case of multiple data files, the program read an additional input parameter file  \texttt{nf\_input\_set.dat} for the additional datafile names  to analyze and data ranges to consider.

The exploration of the parameter space and the corresponding evaluation of the likelihood function is done implementing the nested sampling algorithm \cite{Skilling2004,Sivia,Skilling2006}.
If the data are in the format $(channel, counts)$, a Poisson distribution for each channel is assumed for the likelihood function.
If the data has error bars $(channel, y, \delta y)$, a Gaussian distribution is assumed (new feature in V.~2.2).

The main analysis results are summarized in the output file \texttt{nf\_output\_res.dat} .
Here the details of the computation (n. of live points, n. of  trials, n. of total iteration) can be found as well as the final evidence value and its uncertainty $E \pm \delta E$, the parameter values $\boldsymbol{\hat a}$ corresponding to the maximum of the likelihood function, the mean, the median, the standard deviation and the confidence intervals (68\%, 95\% and 99\%) of the posterior probability distribution of each parameter.
The information gain $\mathcal{H}$ and the Bayesian complexity $\mathcal{C}$ are also provided in the output.

Data for plots and for further analyses are provided in the files \texttt{nf\_output\_data\_*.dat}.
These files contain the original input data together with the model function values corresponding to the parameters  with the highest likelihood function value (\texttt{nf\_output\_data\_max.dat}) or the parameter mean value (\texttt{nf\_output\_data\_mean.dat})  or  median value  (\texttt{nf\_output\_data\_median.dat}) with the corresponding residuals and error bars.
Additional \texttt{nf\_output\_fit\_*.dat} files contain a model evaluation with higher density than the original data for graphical presentation purpose.

The step-by-step details of the nested sampling exploration are provided in the file \texttt{nf\_output\_points.dat} that contains the live points used during the parameter space exploration, their associated  likelihood values and posterior probabilities.
From this file, the different parameter probability distributions and joint probabilities can be built from the marginalization of the unretained parameters. 
For this purpose, a special dedicated Phython library \texttt{Nested\_res} has been developed.
Additional informations can be found in Ref.~\cite{Trassinelli2017b}.

\section{Implementation of the nested sampling for the evidence calculation} \label{sec:nested}

For a given data set(s) $ \{x_i, y_i\}$  and model(s) $\mathcal{M}$, the Bayesian evidence $P( \{x_i, y_i\} | \mathcal{M}, I)$  is extracted for the evaluation of the probability to the different models them-selves:
\begin{equation}
P(\mathcal{M} | \{x_i, y_i\} ,I) \propto P( \{x_i, y_i\} | \mathcal{M}, I) \times P(\mathcal{M} | I), \label{eq:bayes-data}
\end{equation}
where $P(\mathcal{M} | I)$ is the prior probability of each model (assumed constant if not specific preferences for the model is present) and $I$ indicates the background information.
The Bayesian evidence is the integral value of the likelihood function over the entire parameter space defined by the priors $P(\boldsymbol{a}| \mathcal{M}, I)$: 
\begin{equation}
E(\mathcal{M}) \equiv P( \{x_i, y_i\} | \mathcal{M}, I) 
=  \int P( \{x_i, y_i\} | \boldsymbol{a},\mathcal{M}, I) P(\boldsymbol{a}| \mathcal{M}, I) d^{J}\boldsymbol{a} 
  = \int L^\mathcal{M}(\boldsymbol{a}) P(\boldsymbol{a}| \mathcal{M}, I) d^{J}\boldsymbol{a}, \label{eq:evidence}
\end{equation}
where $J$ is the number of the parameters of the considered model, and where we  explicitly show the dependency of likelihood function $L^\mathcal{M}(\boldsymbol{a})$ on the model $\mathcal{M}$. 

The calculation of the Bayesian evidence is made with the nested sampling, similarly to other available codes \cite{Sivia, Mukherjee2006,Feroz2008,Feroz2009,Veitch2010}.
Nested sampling allows for reducing the above integral in the one-dimensional integral
\begin{equation}
E(\mathcal{M}) = \int _0^1 \mathcal{L}(X) d X, \label{eq:evX}
\end{equation} 
where $X$ is defined by the relation
\begin{equation}
X(\mathcal{L}) = \int_{L(\boldsymbol{a})>\mathcal{L}} P(\boldsymbol{a}| I) d^{J}\boldsymbol{a}. \label{eq:XvsL}
\end{equation}
Eq.~\eqref{eq:evX} can be numerically calculated using the rectangle integration method subdividing the $[0,1]$ interval in $M+1$ segments with an ensemble $\{X_m\}$ of $M$ ordered points 
$0< X_M<...<X_2<X_1<X_0=1$. 
We have then 
\begin{equation}
E(\mathcal{M}) \approx \sum_m \mathcal{L}_m \Delta X_m, \label{eq:evindence_app}
\end{equation} 
where $ \mathcal{L}_m = \mathcal{L}(X_m)$ and $\Delta X_m= X_m - X_{m+1}$.
The evaluation of $\mathcal{L}_m$ is obtained by the exploration of the likelihood function with a Monte Carlo sampling via a subsequence of steps.
For this, we use a collection of $K$ parameter values $\{\boldsymbol{a}_k\}$ that we call \textit{live points}. 
More details on the nested sampling algorithm and its implementation can be found in Refs.~\cite{Skilling2004,Skilling2006,Sivia,Mukherjee2006,Feroz2008,Feroz2009,Veitch2010}.
The specific implementation of nested sampling in \texttt{Nested\_fit} is presented in details in Ref.~\cite{Trassinelli2017b}.

The bottleneck of the nested sampling algorithm is the search of new points within the $J$-dimensional volume defined by $L>\mathcal{L}_m$.
Different methods are commonly employed to accomplish this difficult task. 
One efficient method is the ellipsoidal nested sampling \cite{Mukherjee2006}. It is based on the approximation of the iso-likelihood contour defined by $L =\mathcal{L}_m$ by a $J$-dimensional ellipsoid calculated from the covariance matrix of the live points.
The new point is then selected within the ellipsoidal volume (with an enlargement factor selected by the user).
This method, well adapted for unimodal posterior distribution has also been extended to multimodal problems \cite{Feroz2008,Feroz2009}, i.e. with the presence of distinguished regions of the parameter space with high values of the likelihood function. 
Other search algorithms are based on Markov chain Monte Carlo (MCMC) methods \cite{Veitch2010} and the recent 
\textit{Galilean Monte Carlo} \cite{Skilling2012,Feroz2013}, particularly adapted to explore the regions close to the boundary of $V_{L>\mathcal{L}_m}$ volumes.
\texttt{Nested\_fit} program is based on an improved version of the \textit{lawn mower robot} method, originally developed by L.~Simons \cite{Theisen2013} and presented in details in the next section.

\section{The lawn mower robot search algorithm}  \label{sec:lawn-mower}

\begin{figure}
\centering
\includegraphics[width=0.45\textwidth]{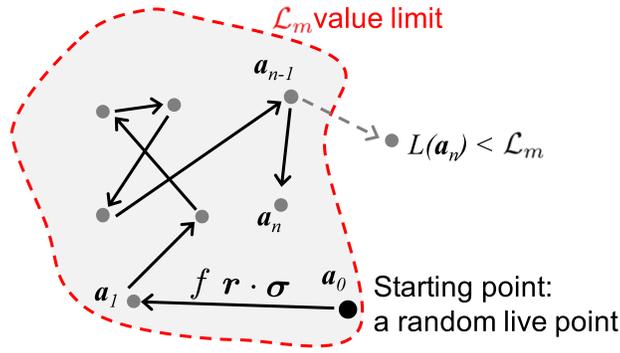}
\caption{Scheme of lawn mower robot algorithm.}
\label{fig:lawn-mower}
\end{figure} 

A schematic view of the improved  \textit{lawn mower robot} algorithm is represented in Fig.~\ref{fig:lawn-mower}.
To cancel the correlation between the starting point and the final point, a series of $N$ jumps are made in this volume.
The different stages of the algorithm are 
\begin{enumerate}
\item Choose randomly a starting point $\boldsymbol{a}_{n=0} = \boldsymbol{a}_0$ from the available live points $\{\boldsymbol{a}_{m,k}\}$ as starting point of the Markov chain where $n$ is the number of the jump. 
The number of tries $n_t$ (see below) is set to zero.

\item From  the values $\boldsymbol{a}_{n-1}$, find a new parameter sets $\boldsymbol{a}_{n}$  where each $j^\text{th}$ parameter is calculated by $(a_{n})_j= (a_{n-1})_j +  f\ r_j \sigma_j$, where $\sigma_j$ is the standard deviation of the live points of the nested sampling computation step relative to the $j^\text{th}$ parameter, $r_j \in [-1,1]$ is a sorted random number and $f$ is a factor defined by the user.
\begin{enumerate}
\item If $L(\boldsymbol{a}_{n})>\mathcal{L}_m$ and $n< N$, go to the beginning of step 2 with an increment of the jump number $n = n + 1$.

\item If $L(\boldsymbol{a}_{n})>\mathcal{L}_m$ and $n= N$, $\boldsymbol{a}_{n=N}$ is new \textit{live point} to be included in the new set $\{\boldsymbol{a}_{m+1,k}\}$.

\item If $L(\boldsymbol{a}_{n})<\mathcal{L}_m$ and $n< N$ and the number of tries $n_t$ is less than the maximum allowed number $N_t$, go back to beginning of step 2 with an increment of the number of tries $n_t = n_t + 1$. \label{step}

\item If $L(\boldsymbol{a}_{n})<\mathcal{L}_m$ and $n< N$ and $n_t = N_t$ a new parameter set $\boldsymbol{a}_0$ has to be selected.
Instead than choosing one of the existing live points, $\boldsymbol{a}_0$ is built from distinct $j^\text{th}$ components from different live points: $(a_0)_j = (a_{m,k})_j$ where $k$ is randomly chosen between 1 and $K$ for each $j$. 
 Then $\boldsymbol{a}_{n=0} = \boldsymbol{a}_0$ and go to the beginning of step 2.

\end{enumerate}

\end{enumerate}

Step \ref{step}, the main improvement of the original lawn mower robot algorithm, makes the algorithm well adapted to problems with multimodal parameter distributions allowing easy jump between high-likelihood regions. 
The value of $N_t$ is fixed in the code ($N_t = 10000$ in the present version). 
The other parameters can be provided by the input file.

\section{An application to low-statistics data} \label{sec:go}

To show the capabilities of \texttt{Nested\_fit}, we present in this section its implementation on a particular critical case corresponding to a debated experiment.
In 2008 it was observed an unexpected modulation in the two-body electron capture decay of single H-like $^{142}_{~61}$Pm ions to the stable $^{142}_{~60}$Nd bare nucleus, with a monochromatic electron-neutrino emission \cite{Litvinov2008}.
The same modulation frequency, but with much smaller amplitude, was found in 2010 data \cite{Kienle2013} but not in the latest campaign in 2014 \cite{Ozturk2019} where much more events have been recorded.

\begin{figure}
\centering
\includegraphics[width=0.65\textwidth]{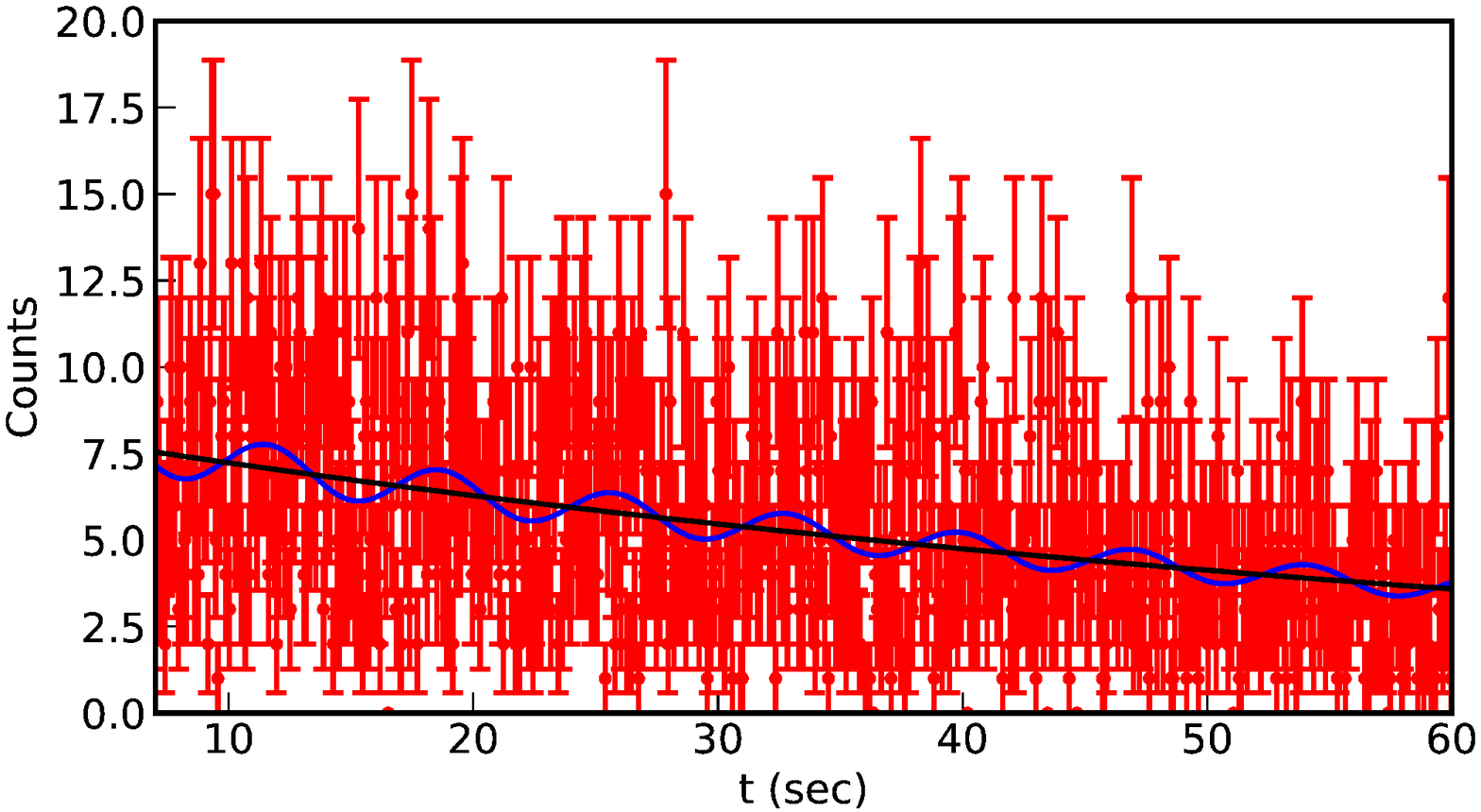}
\includegraphics[width=0.65\textwidth]{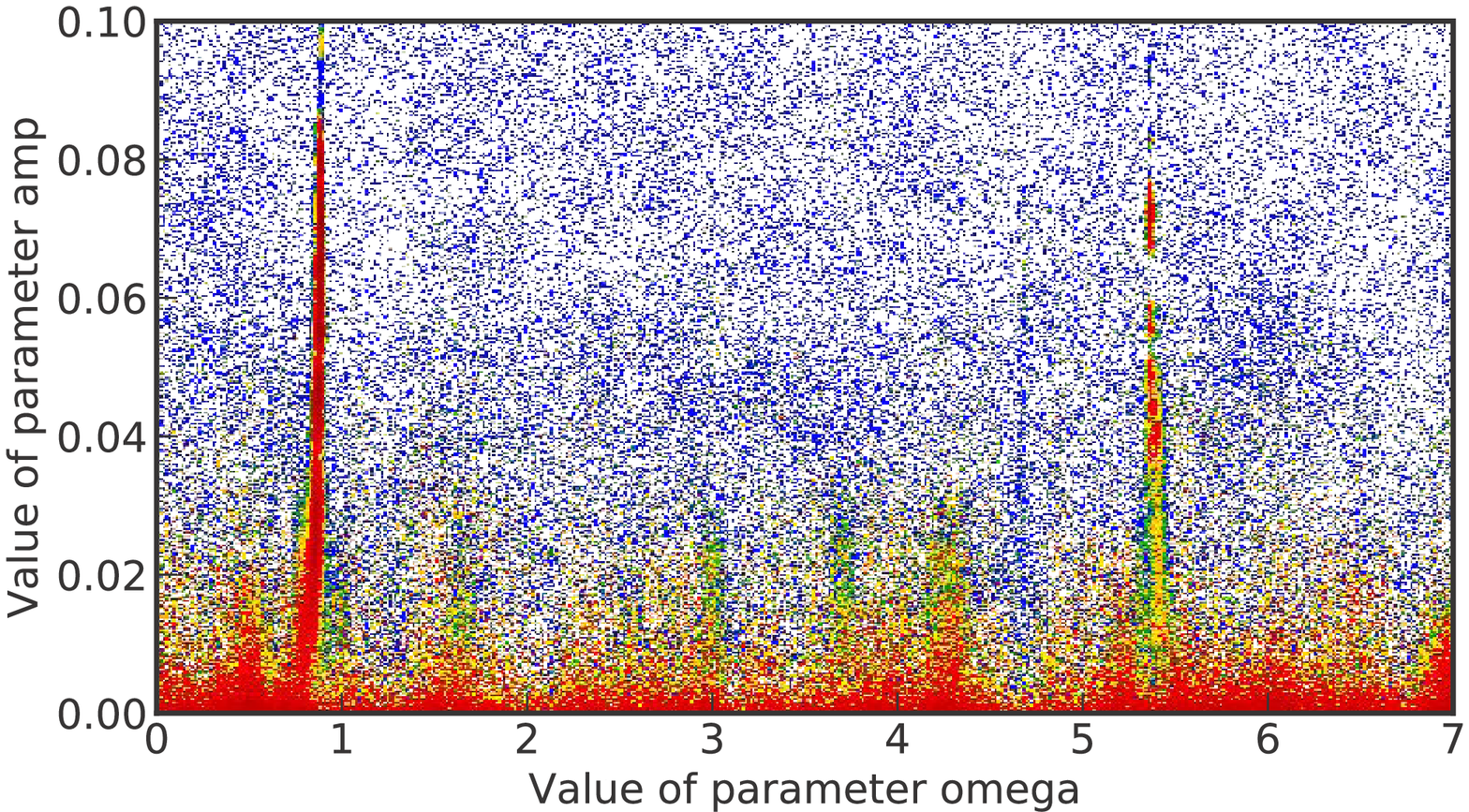}
\caption{Top: Data relative to the single decay of H-like $^{142}_{~61}$Pm to $^{142} _{~60}$Nd bare nucleus obtained with a binning of 0.08~s.
The profile curves relative to pure exponential and exponential with modulation models are also represented.
Bottom: 2D histogram of the joint probability of the amplitude $a$ and pulsation $\omega$ of the model with modulation.
Red, yellow and green colors represent approximatively the regions corresponding to 68\%, 95\% and 66\% confidence intervals.
Both figures are obtained by Python  \texttt{nested\_res.py} library that accompany  \texttt{Nested\_fit} program.}
\label{fig:go}
\end{figure}

The unstable ions are produced by collision with a solid target and then injected in a storage ring where they are cooled down.
In the storage ring, the decay time of single ions is measured from changes of the Schottky noise frequency induced by the ion revolution.
The H-like $^{142}_{~61}$Pm ion and $^{142}_{~60}$Nd bare nucleus  masses correspond in fact to different revolution frequencies.
From the accumulated data of single decay events, the decay probability per unit of time can be measured.
An example of the data collected in 2010 is presented in Fig.~\ref{fig:go} (up).

The observed modulation of the expected exponential decay has not yet a clear explanation. A possible connection with neutrino masses differences is speculated in the literature.
The determination of the presence or not of a modulation is a perfect case for implementing Bayesian model comparison with \texttt{Nested\_fit}.

When a possible modulation of the exponential decay is assumed,  the likelihood function corresponding to 2010 data presents several maxima.
This reflect the periodicity nature of the considered function, which can manifest itself via different harmonics, and the low number of available counts per channel.
The difficulties to deal with these multiple likelihood maxima pushed in fact the creation of the improved \textit{lawn mower robot} algorithm. 

\begin{table}[H]
\caption{Summary of the results provided from \texttt{Nested\_fit} for the two considered models.
The parameter values are given in terms of most probable value and 95\% confidential interval (CI).}
\centering
\begin{tabular}{lcc}
\toprule
& \textbf{Model 1}	& \textbf{Model 2}	\\
\midrule
Function & $y = N_0 e^{- t / \tau}$ & $y = N_0 e^{- t / \tau} [1+ a \sin (\omega t + \phi)]$ \\
\midrule
log$_e$(Evidence) &  $-1594.11 \pm 0.30$	& $-1594.60 \pm 0.36$		\\
Probability & 34.2--41.9\% 	& 58.1--68.8\%		         \\
Complexity & 2.05		& 15.19		         \\
Extracted information [nat] &  4.76 &  6.32 \\ 
$\omega$ (CI 95\%) [rad s$^{-1}$] &  -- &  $0.89 (0.17-6.86)$ \\
$a$ (CI\ 95\%) &  --  &  $9.2\times10^{-2} (2.2\times10^{-4}-7.2\times10^{-2})$ \\
$\phi$ (CI\ 95\%) [rad] & -- & $3.84(0.18-6.14)$ \\
\bottomrule \label{tab}
\end{tabular}
\end{table}

In Fig.~\ref{fig:go} (top) we present the collected data together with the exponential and modulated exponential functions corresponding to the most probable parameter set.
The output result from \texttt{Nested\_fit} are presented in Tab.~\ref{tab} where model 1 and 2 represent the absence of presence of modulation.
For each model, values of the evidence, Bayesian complexity and extracted information are provided, as well as model probabilities. 
The uncertainty of the probabilities is related to the uncertainty of the evidence.
As example of probability distribution, we present  in Fig.~\ref{fig:go} (bottom) the joint probability of the amplitude $a$ and pulsation $\omega$  of the modulation in model 2. 
The 2D histogram (obtained with Python \texttt{nested\_res.py} library that accompany \texttt{Nested\_fit} program) is constructed by marginalization on the other parameters. 
As it can be seen, different maxima are visible, which make difficult the convergence of the nested sampling method. 
The improved \textit{lawn mower robot} algorithm can deal with this kind of situation, even if the computation time is sometime long (several days in a single CPU). 

As it can be observed, the assigned probability to each model are similar and the confidential intervals for the parameter relative to the modulation model are very large.
These two aspects reflect the difficulty to treat this problem where the acquired data are not sufficient to provide a marked preference for one model with or without modulation (see Ref.~\cite{Ozturk2019} for a more extended discussion).
Even if apparently unsatisfying, this result avoid however possible over-interpretation of the data commonly encountered when classical methods are employed, as recently discussed in Ref.~\cite{King2019} in the context of nuclear physics.

\section{Conclusions}
We presented here the program \texttt{Nested\_fit}, a general purpose parallelized data analysis code for the evaluation of Bayesian evidence and other statistically relevant outputs.
It uses the nested sampling method with the implementation of the improved lawn mower robot algorithm for the evaluation of the Bayesian evidence.
\texttt{Nested\_fit} has been developed over the past years for the analysis of several sets of atomic  experimental data that strongly contribute to the code evolution.
We would like to mention in particular the analysis of low-statistics X-ray spectra of He-like uranium \cite{Trassinelli2009,Trassinelli2017b}, X-ray spectra of pionic atoms \cite{Trassinelli2016b,Trassinelli2016c}, electron photoemission spectra from nano-particles \cite{Papagiannouli2018,DeAndaVilla2019}, single-ion decay spectra \cite{Ozturk2019} and response function of crystal X-ray spectrometers (in progress).

Compared to the version reported in Ref.~\cite{Trassinelli2017b},  the presented version (V. 2.2) shows additional important features:  i) the possibility to interpolate and use computed or simulated external profiles and ii) the implementation of Gaussian likelihood function for data with error bars.

Future developments of \texttt{Nested\_fit} will be focussed on the implementation of new exploration methods for the live point evolution of the nested sampling \cite{Feroz2008,Feroz2009,Skilling2012,Feroz2013}.
More precisely, the main goal is the improvement the efficiency for the exploration of the parameter space where the likelihood function presents several local maxima.

\vspace{6pt} 



\funding{This research received no external funding.}

\acknowledgments{The author thanks the Alexander von Humboldt Foundation that provide the financing to attend the MaxEnt2019 conference.
Moreover, the author would like to express once again his deep gratitude to Leopold M. Simons who introduced the author to the Bayesian data analysis and without whom this work could not have been started. 
The development of this program would not be possible without the close interactions and discussions with many collaborators that the author would like to thank as well: N. Winckler, R. Grisenti, A. Lévy, D. Gotta, Y. Litvinov, J. Machado, N. Paul and all members of the Pionic Hydrogen, FOCAL and GSI Oscillation collaborations and the ASUR group at INSP.}

\conflictsofinterest{The authors declare no conflict of interest.} 



\reftitle{References}

\end{document}